\documentclass[12pt]{article}

\usepackage{aasms4,natbib,tighten}

\def\ifundefined#1{\expandafter\ifx\csname#1\endcsname\relax}
\def\la{\mathrel{\hbox{\rlap{\hbox{\lower4pt\hbox{$\sim$}}}\hbox{$<$}}}}
\def\ga{\mathrel{\hbox{\rlap{\hbox{\lower4pt\hbox{$\sim$}}}\hbox{$>$}}}}
\newcommand{\be}{\begin{eqnarray}}
\newcommand{\ee}{\end{eqnarray}}
\ifundefined{nuc}\def\nuc#1#2{\relax\ifmmode{}^{#1}{\protect\text{#2}}\else${}^{#1}$#2\fi}\else\relax\fi
\ifundefined{ensuremath}\def\ensuremath#1{\relax\ifmmode{#1}}
\else\hbox{${#1}$}\fi\else\relax\fi

\newcommand{\etal}{et al.}

\newcommand{\kmps}{km~s$^{-1}$}

\newcommand{\msol}{\ifmmode{{\rm M}_\odot}\else{M$_\odot$}\fi}
\newcommand{\foe}{\ifmmode{10^{51}}\else{$10^{51}$}\fi}
\newcommand{\nni}{\nuc{56}{Ni}}
\newcommand{\xni}{\ifmmode{{\rm X}_{\rm Ni}}\else{X$_{\rm Ni}$}\fi}
\def\ang{\hbox{\AA}}

\def\tstd{\ifmmode{\tau_{\rm std}}\else{\hbox{$\tau_{\rm std}$} }\fi}
\def\Rzero{\ifmmode{R_0}\else{\hbox{$R_0$} }\fi}
\newcommand{\vno}{\ifmmode{v_0}\else{\hbox{$v_0$} }\fi}
\def\tdecay{\ifmmode{t_{\rm decay}}\else{\hbox{$t_{\rm decay}$} }\fi}
\begin{document}

\bibliographystyle{natbib-apj}

\title{Synthetic Spectra of Hydrodynamic Models of Type
Ia Supernovae} \author{Peter Nugent\altaffilmark{1,2},
E.~Baron\altaffilmark{2}, David Branch\altaffilmark{2}, Adam
Fisher\altaffilmark{2} and Peter H. Hauschildt\altaffilmark{3,4}}

\altaffiltext{1}{Lawrence Berkeley Laboratory, University of
California, 1 Cyclotron Road, Mail Stop 50-232, Berkeley, CA, 94720;
penugent@lbl.gov} 

\altaffiltext{2}{Dept. of Physics and Astronomy, University of
Oklahoma, 440 W.  Brooks, Rm 131, Norman, OK 73019-0225;
baron,branch,fisher@phyast.nhn.ou.edu}

\altaffiltext{3}{Dept. of Physics and Astronomy, 
University of Georgia, Athens, GA 30602-2451; yeti@hal.physast.uga.edu}

\altaffiltext{4}{Dept. of Physics and Astronomy, Arizona State
University, Tempe, AZ 85287-1504}

\begin{abstract}
We present detailed NLTE synthetic spectra of hydrodynamic SNe Ia
models. We make no assumptions about the form of the spectrum at the
inner boundary. We calculate both Chandrasekhar-mass deflagration
models and sub-Chandrasekhar ``helium detonators.'' Gamma-ray
deposition is handled in a simple, accurate manner.  We have
parameterized the storage of energy  that arises from the time dependent
deposition of radioactive decay energy
in a reasonable manner, that spans the expected range. We find that
the Chandrasekhar-mass deflagration model W7 of Nomoto \etal\ shows
good agreement with the observed spectra of SN~1992A and
SN~1994D, particularly in the UV, where our models are expected to be
most accurate. The sub-Chandrasekhar models do not reproduce the UV
deficit observed in normal SNe~Ia. They do bear some resemblance to
sub-luminous SNe~Ia, but the shape of the spectra (i.e. the colors)
are opposite  to that of the observed ones and the intermediate mass element
lines such as \ion{Si}{2}, and \ion{Ca}{2} are extremely weak, which
seems to be a generic difficulty of the models. Although the
sub-Chandrasekhar models have a significant helium abundance (unlike
Chandrasekhar-mass models), helium lines are not prominent in the
spectra near maximum light and thus do not act as a spectral signature
for the progenitor.
\end{abstract}


\section{Introduction}

Type Ia supernovae (SNe Ia) are among the brightest known objects in
the universe. Since they form a nearly homogeneous class and simple
selection criteria can make the observed dispersion quite small, they
are natural cosmological probes \cite[]{vauetal95}. The observed
homogeneity has led to a search for a homogeneous progenitor, that
would satisfy the requirement of lacking hydrogen. This has led to the
assumption that the SNe~Ia progenitor involves the explosion of a
Chandrasekhar-mass white dwarf. The current status of the search for
the identification of the SNe~Ia progenitor is reviewed in
\cite{prog95}. Hydrodynamic explosion models have included
deflagration models such as the ``W7'' model of \cite{nomw7}. While
this model is somewhat hand-crafted to fit the observed spectra and
suffers from an overproduction of neutron rich species, it remains the
standard in the field. The ``DD'' \cite[delayed
detonation,][]{khok91a,woosdd91} and ``PDD'' models \cite[Pulsating
Delayed Detonation,][]{khok91b} improve the predicted nucleosynthetic
yield and gives qualitative agreement with the observed spectra and
light curves \cite[]{hofkhoklc96}.

The observation of the super-luminous SN~1991T \cite[]{jeffetal92} and
the very sub-luminous SN~1991bg \cite[]{fil91bg92} convincingly showed
that the class of SNe~Ia is not entirely homogeneous. In attempting to
model SNe~Ib, \cite{livglas90,livglas91} examined a ``helium-igniter''
where a sub-Chandrasekhar or a Chandrasekhar-mass C/O white dwarf with
an accreted helium shell detonates near the center following the
detonation of the helium shell.  Motivated by observations of SN 1991T
and 1991bg as well as other claims for evidence that SNe~Ia form a
sequence \cite[]{philm15,hametal95} \cite{wwsubc94} and \cite{livar95}
investigated the helium-igniter as a realistic model for SNe~Ia.

On the face of it helium-igniters have much to recommend them as a
plausible progenitor model of SNe~Ia. By varying the initial white
dwarf mass such models can produce a range of nickel mass $M_{\rm Ni}
\approx 0.2-1.0$~\msol, naturally leading to a sequence of supernovae.
Sub-Chandrasekhar models do not suffer from the same neutronization
problem that occurs in the C/O deflagration models, and population
synthesis studies may produce the progenitors in the requisite
quantities \cite[]{tyi92}, although not in old populations
\cite[]{prog95}.

On the negative side, the models produce an outer shell of the
products of explosive helium burning: helium and \nni; elements not
typically associated with the outer layers of SNe~Ia. In addition the
light curves are extremely fast \cite[]{wwsubc94,hofkhoklc96}, so in
particular, it is not clear that the observed photometric diversity
can be reproduced by such models.

Since one of the primary goals of synthetic spectral synthesis is the
confrontation of theoretical models with observations, we present the
results of synthetic spectrum calculations of the helium igniter
models of \cite{wwsubc94} and \cite{livar95}. Since this is our first
application of our program of synthetic spectral synthesis to
hydrodynamical SNe~Ia models, we also present the results of the spectral
synthesis of the W7 model \cite[]{nomw7}.

\section{Calculations}

The calculations are performed using the generalized stellar
atmosphere program {\tt PHOENIX 7.1}
\cite[]{phhs392,phhre92,phhcas93,hbapara96} in generally the same way
we have applied it previously to SNe~Ia \cite[]{nug1a95,nugseq95},
although we have modified the code to allow the treatment of nebular
boundary conditions, as well as stratified composition and a full
gamma-ray deposition calculation. We have compared the results of our
$\gamma$-ray deposition with more detailed calculations and the
agreement is excellent (Young \& Kumagai 1995, private
communication). The boundary conditions make no assumptions about the
form of the flux at the inner boundary, but rather impose continuity
requirements on the intensity (with correct Lorentz
transformations). Thus all of the flux comes from the atmosphere
itself, there is no ``light bulb'' at the center. {\tt PHOENIX}
accurately solves the fully relativistic radiation transport equation
along with the non-LTE rate equations (for some ions) while ensuring
radiative equilibrium (energy conservation). The following ions were
treated in non-LTE in the calculations reported here (the number of
levels follows in parenthesis): \ion{He}{1} (11), \ion{He}{2} (10),
\ion{Na}{1} (3), 
\ion{Ne}{1} (26), \ion{Ca}{2} (87), \ion{Mg}{2} (18), \ion{C}{1} (228),
\ion{O}{1} (36), 
\ion{Fe}{2} (617), \ion{Co}{2} (255), \ion{Ti}{2} (204), \ion{S}{2}
(85) and \ion{Si}{2} (94).

The hydrodynamical models were evolved in time by assuming that the
expansion is homologous, i.e. the velocity of any given mass point was
held constant. The models were rezoned into 50 mass zones, with
roughly a logarithmic spacing in \tstd\unskip, where \tstd is the total
extinction optical depth in the continuum at $5000$~\ang. Care was
taken to resolve the density profiles.

In theory, once we choose a time since explosion the model is
completely determined. The density structure is specified by the
homology transformation, the compositions are fixed (once decay of the
radioactive species has been accounted for) and, since we use observed
bolometric luminosity as an input parameter, the temperature structure
is then completely determined by imposing the condition of radiative
equilibrium. We parameterize the luminosity as:
\[ L_{\rm bol} = \eta L^{\rm abs}_\gamma, \]
where $L^{\rm abs}_\gamma$ is the total instantaneous $\gamma$-ray
luminosity {\em deposited in the material} and $\eta$ is a parameter
that measures the net amount of energy stored over time by the
material. Note that $\eta$ differs from the parameter $\alpha$ defined
by Arnett and co-workers \cite[]{arnett82,abw85} since $\alpha$ refers
to the {\em total} instantaneous $\gamma$-ray luminosity, and instead
corresponds to the parameter $\tilde Q$ of \cite{hofkhoklc96}. [NB:
While the definition of $Q\equiv \alpha$ in \cite{hofkhoklc96}, in
previous papers in their series $Q$ corresponds to
$\eta$]. \cite{hofkhoklc96} found $\eta$ in the range $0.7 < \eta <
1.8$ for a wide variety of models that they examined, and we have
varied $\eta$ from approximately $0.5-2.0$. Actually $\eta$ should be
a function of radius, but an accurate calculation of $\eta$ will
require a NLTE, multi-group radiation-hydrodynamical calculation.
This procedure accounts for the time-dependent nature of the
deposition of radioactive energy in an accurate manner.
Given an input luminosity the temperature structure of the models is
determined by demanding the modified radiative equilibrium condition:
\[ \int \kappa_\lambda (B_\lambda - J_\lambda) \, d\lambda - \dot S =
0, \] 
where $\dot S$ is the local instantaneous rate at which
$\gamma$-ray energy is deposited.

\section{Results}

\subsection{Model W7}

\begin{figure}
\epsscale{0.8}
\plotone{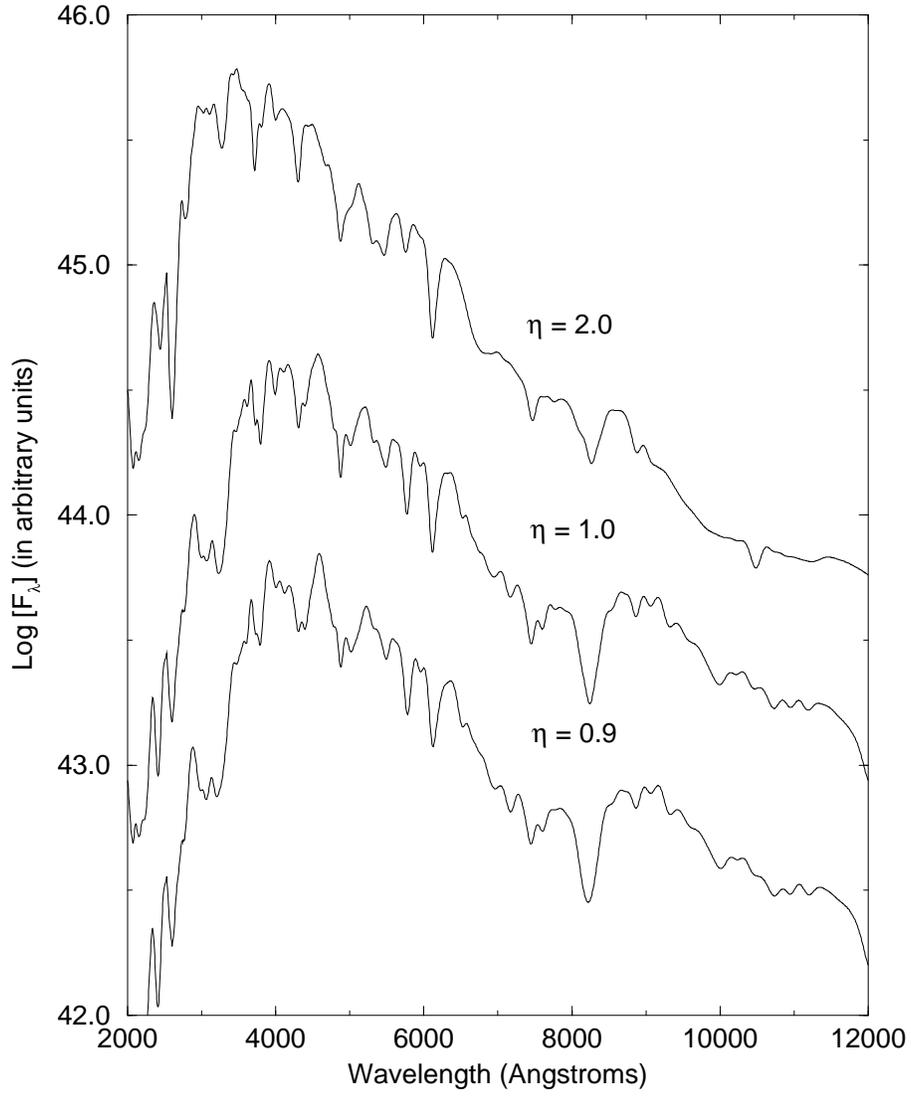}
\caption{\label{w720d} The synthetic spectrum for the W7 model at 20~d
past explosion for three choices of $\eta$.}
\end{figure}

Figure~\ref{w720d} displays our synthetic spectrum for the W7 model at
20~d past explosion for three choices of $\eta$. This is several days
after the time of bolometric maximum $t_{\rm bol} = 14$~d found by
\cite{hofkhoklc96} and \cite{kmh93} who found $\eta = 1.3$ at this
time. The magnitudes and colors of these models are listed in
Table~\ref{tab1}. The model strongly resembles observed SNe~Ia
spectra, showing the defining \ion{Si}{2} $\lambda 6355$ line as well
as the $\lambda 5972$ line, and lines from \ion{Ca}{2}, \ion{S}{2},
\ion{O}{1}, and \ion{Fe}{2}. Also, there is the strong UV deficit that
is characteristic of SNe~Ia spectra. The colors for the $\eta=1.0$
20~d model are very similar (on average) to the colors found for
normal SNe~Ia. The $\eta=2.0$ colors are too blue and the $\eta=0.8$
are too red, suggesting that the value of $\eta=1.3$, found by
H\"oflich and collaborators is reasonable.

\begin{figure}
\epsscale{0.8}
\plotone{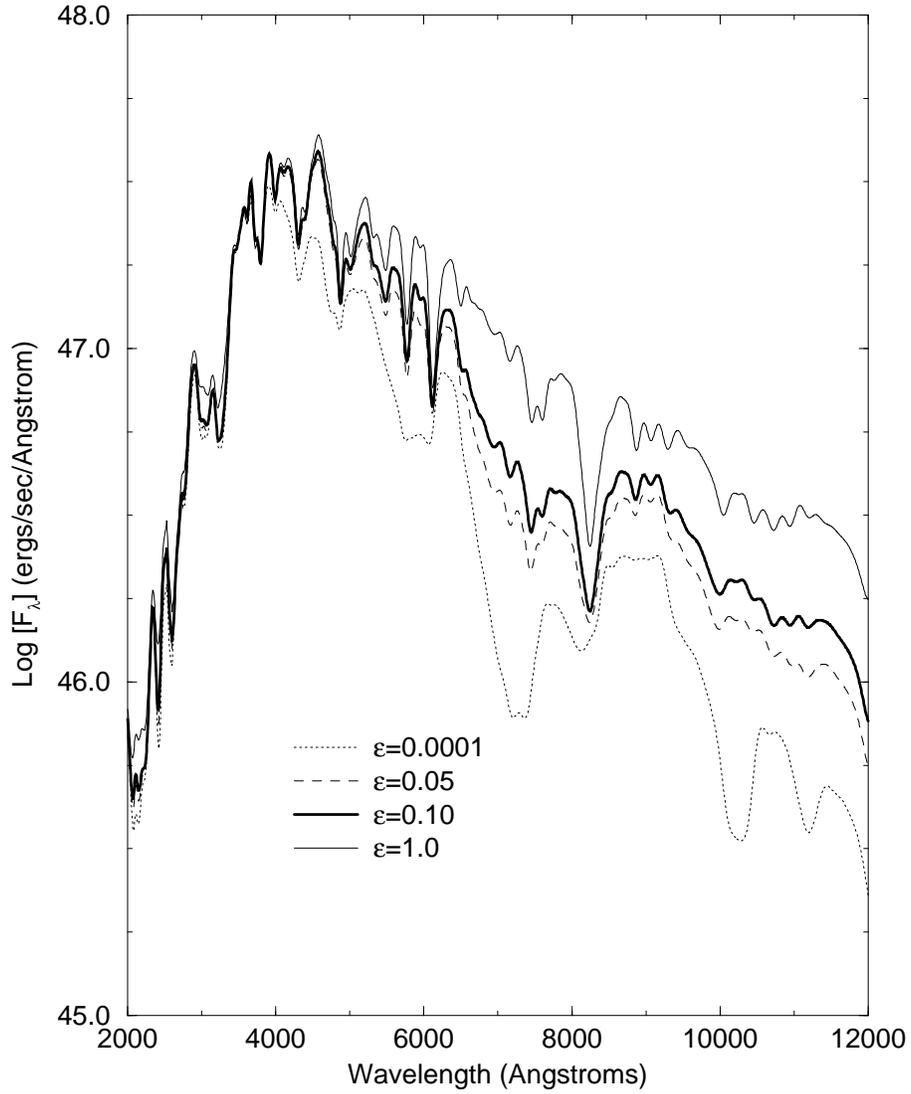}
\caption{\label{nlte}The synthetic spectrum for the W7 model at 23~d
past explosion ($\eta = 1.3$) for four choices of $\epsilon$. The
model with $\epsilon = 0.10$ is in radiative equilibrium and the
temperature structure has been held fixed at that structure for all
the models displayed.}
\end{figure}

In addition to the NLTE lines that we treat directly, we must also include
$\approx 2$~million additional lines in LTE.  The shape of the
spectrum is somewhat sensitive to the constant thermalization
parameter $\epsilon$ that we choose \cite[]{nug1a95,snefe296}, where
$\epsilon$ is defined by the source function for LTE metal lines,
 \[ S_l = (1 - \epsilon)\displaystyle\strut\int\phi_\nu J_\nu
d\nu + \epsilon B_\nu (T). \] Figure~\ref{nlte} compares the spectra
for $\epsilon = (10^{-4},0.05,0.1,1.0)$. The UV (where most lines
are treated in NLTE) is rather insensitive to the choice of
$\epsilon$; however, in the optical, redward of 5000~\ang\ there is a
strong dependence on $\epsilon$. Based on our previous work and the
results of NLTE calculations, we choose
$\epsilon = 0.05 - 0.1$ as our standard range. All the models
discussed below have $\epsilon = 0.1$.

\begin{figure}
\epsscale{0.8}
\plotone{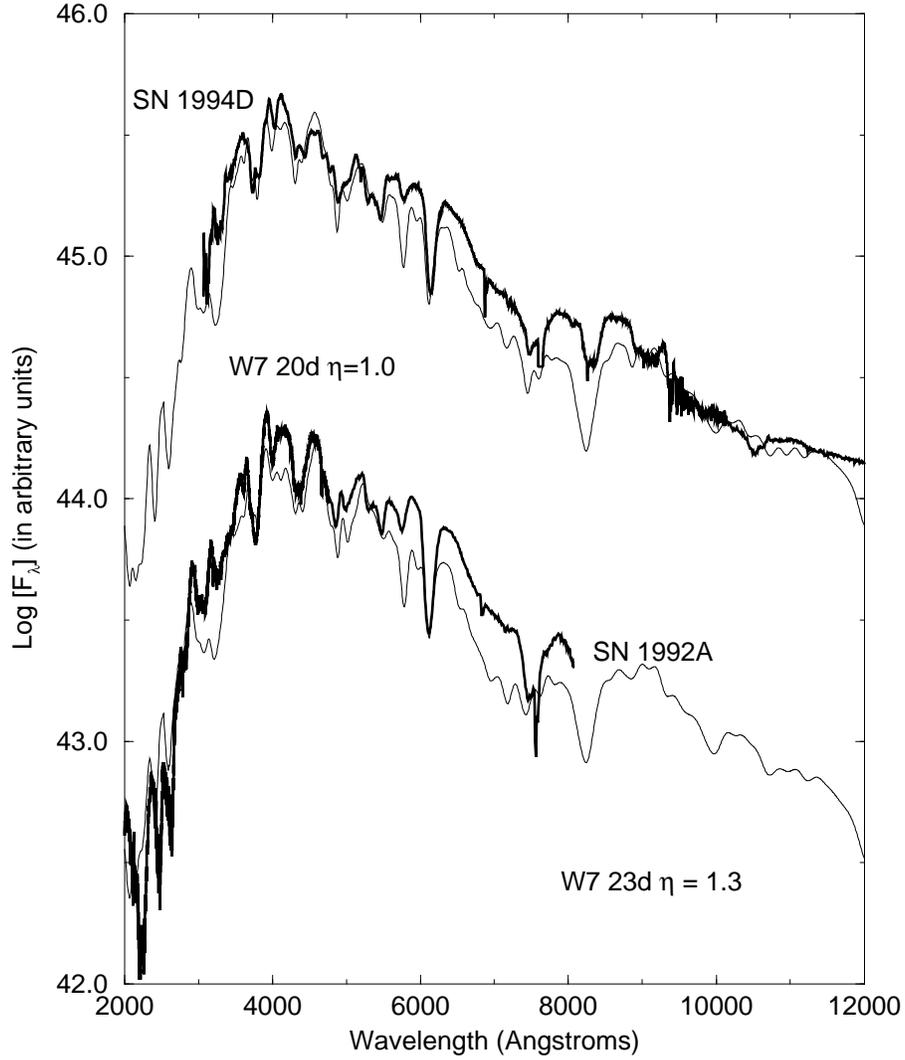}
\caption{\label{comp}The W7 model at 23~d ($\eta=1.3$) compared with
an observed spectrum of SN~1992A [5 days after maximum light
\protect\cite[]{kir92a}] and at 20~d ($\eta=1.0$) compared with an
observed spectrum of SN~1994D [at maximum light
\protect\cite[]{meik94d96}]. }
\end{figure}

Figure~\ref{comp} shows the W7 model at 23~d ($\eta=1.3$) with an
observed spectrum of SN~1992A [taken at 5 days after maximum light
\cite[]{kir92a}] and at 20~d ($\eta=1.0$) with an observed spectrum of
SN~1994D [at maximum light in the optical and 3 days before maximum in the IR
\cite[]{meik94d96}]. The agreement is quite good across the entire
range of observed wavelength for each supernova with all of the major
(and most of the minor) features present in the synthetic
spectra. While fine tuning could no doubt improve the fits, that is
not our purpose in this paper. An interesting feature in
Figure~\ref{comp}, is that both the observed spectrum for SN~1994D and
the synthetic spectra of W7 show a ``split'' just blueward of the
\ion{Ca}{2}~H\&K feature. This is likely due to a blend of
\ion{Ca}{2}~H\&K and \ion{Si}{2} $\lambda 3858$. \cite{kir92a} also noted
that the two lines are of nearly equal strength.  While the split is
prominent in the observed spectrum of SN~1994D, it is clearly absent
in the observed spectrum of SN~1992A.  
We will return to this issue in future work.

\begin{figure}
\epsscale{0.8}
\plotone{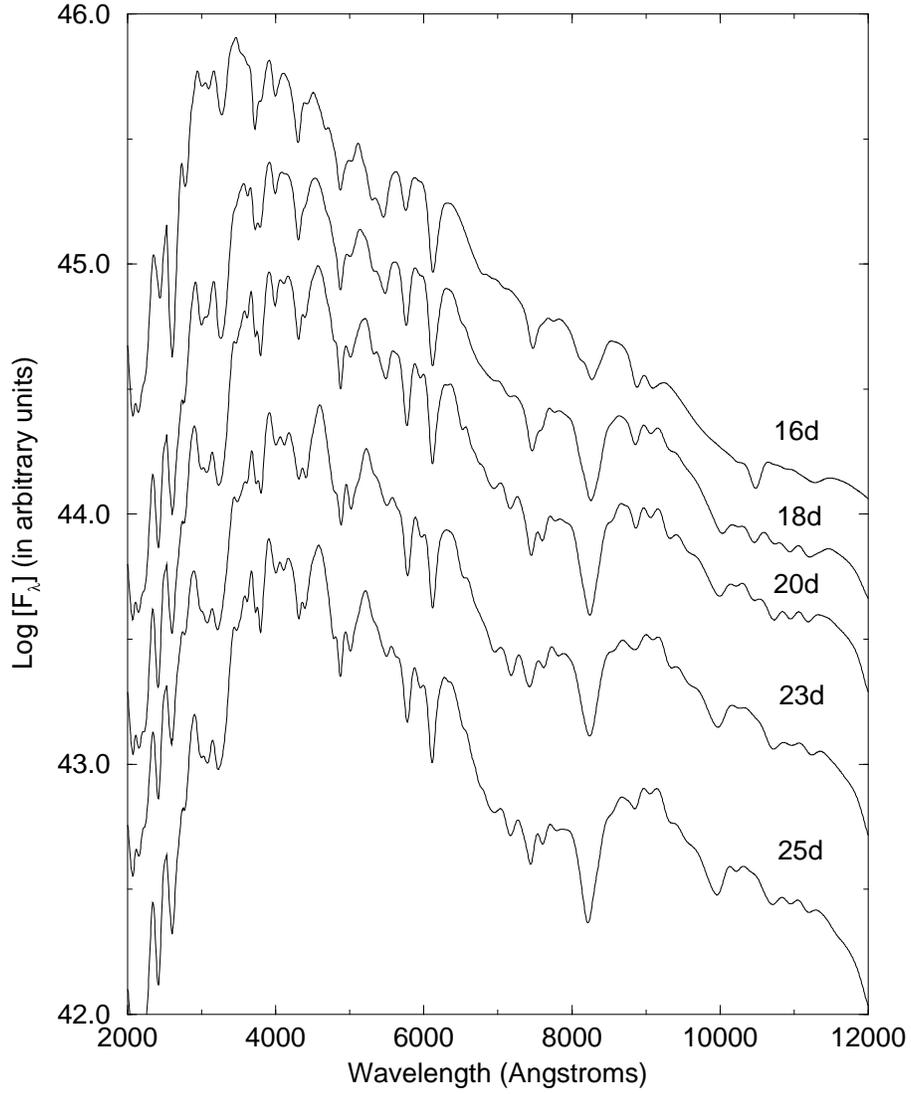}
\caption{\label{w7timev}The time evolution of the W7 model near
maximum light ($\eta=1.1,0.9,1.0,1.3,1.5$ at $t=16,18,20,23,25$~d,
respectively). } 
\end{figure}

Figure~\ref{w7timev} shows the time evolution of the W7 model near
maximum light. The bolometric magnitudes and colors of the 16~d,
$\eta=1.1$ and $\eta=1.7$ models (see Table~\ref{tab1}) should be
compared with the results of \cite{hofkhoklc96} who found $M_{\rm bol}
= -19.56$ and $B-V = 0.11$ (for $\eta = 1.3$). The 
bolometric magnitudes bracket the results of \cite{hofkhoklc96}, while
both of the models are 
somewhat bluer than they found.
This is likely due to differences in the treatment of radiation
transport (NLTE, 82,000 wavelength points and full line profiles,
versus LTE grey transport) and serves as an estimate of the
theoretical uncertainty of such calculations.

\subsection{Sub-Chandrasekhar Models}

\begin{figure}
\epsscale{0.8}
\plotone{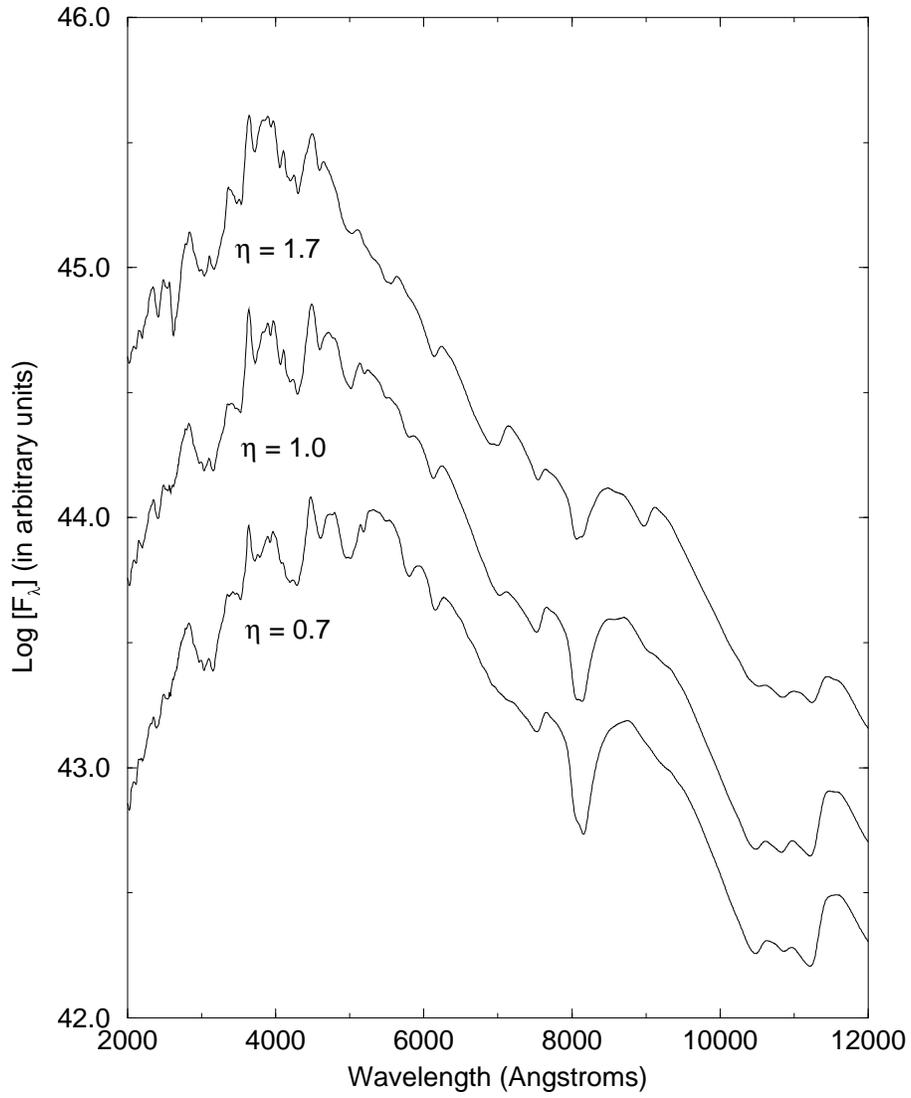}
\caption{\label{woos1}Model WW2 15~d after explosion for three choices
of $\eta$.}
\end{figure}

\begin{figure}
\epsscale{0.8}
\plotone{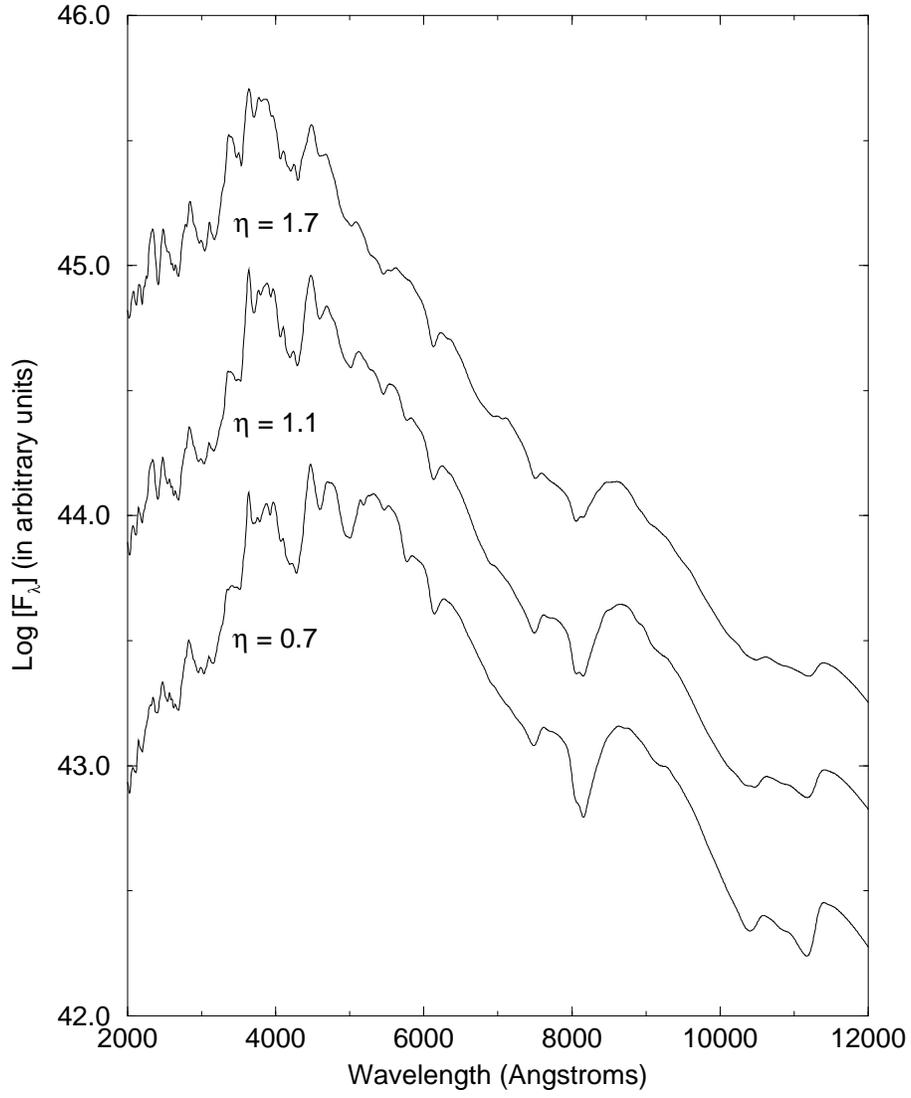}
\caption{\label{arn1}Model LA4 15~d after explosion for three
choices of $\eta$.}
\end{figure}

Figure~\ref{woos1} displays Model 2 of \cite{wwsubc94} (WW2) 15~d
after explosion for three choices of $\eta$. This model is the
explosion of an 0.7~\msol\ white dwarf that has accreted 0.2~\msol\ of
helium. Since these models are significantly less massive than W7,
they peak earlier and hence it is sensible to examine them at earlier
times.  Figure~\ref{arn1} is similar to Figure~\ref{woos1}, but for
Model 4 of \cite{livar95} (LA4), which is a 0.7~\msol\ white dwarf
that has accreted 0.17~\msol\ of helium. The magnitude and color data
for these models can be found in Table~\ref{tab2}. The synthetic
spectra have less line-blanketing and hence more flux in the UV, than
does the W7 model. There is no strong evidence of either \ion{He}{1}
or \ion{He}{2} lines. Although we do not use the Sobolev approximation
at all in our calculations, we calculate the Sobolev optical depth of
each NLTE line as a convenient diagnostic. The Sobolev optical depth
of the \ion{He}{1} $\lambda5876$ lines is approximately $3-4$ orders
of magnitude weaker than that of the \ion{Si}{2} $\lambda6355$ line in
both sets of models, thus these models are effective at ``hiding
helium.'' It is somewhat surprising that in a model with nearly
0.2~\msol\ helium on the outside that no evidence of helium should
appear. This seems to be due to the very strong non-thermal ionization and
the high UV flux which tends to keep  the helium ionized, and/or highly
excited, suppressing the strong
\ion{He}{1} lines. Optical \ion{He}{2} lines, particularly $P_\alpha\
\lambda4687.8 $ and $P_\beta\ \lambda3204.5$, are also not
prominent. At earlier times, with higher densities recombination may
populate the \ion{He}{1}~--~II levels, but since the models will also
be hotter at those times, it is not clear {\em a priori} that optical
\ion{He}{1}~--~II lines will ever be strong in these
models. Understanding the exact suppression mechanism of helium lines
will be the subject of future work.

\begin{figure}
\epsscale{0.8}
\plotone{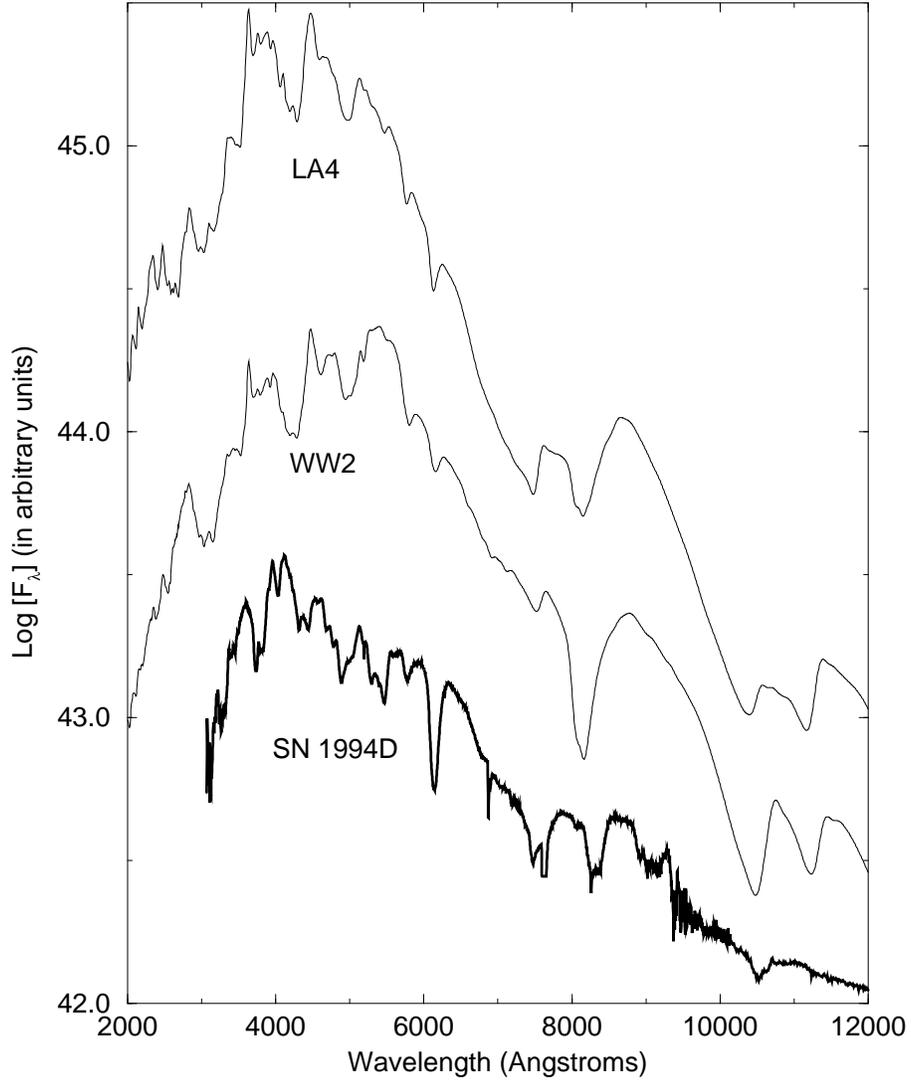}
\caption{\label{subc_comp1}The synthetic spectra of WW2 ($\eta=1.1$)
and LA4 ($\eta=1.5$) at 20~d compared with the observed spectrum of SN
1992A [5 days after maximum light \protect\cite[]{kir92a}].}
\end{figure}

Figure~\ref{subc_comp1} shows the synthetic spectra of the models of
WW2 ($\eta=1.1$) and LA4 ($\eta=1.5$) at 20~d, compared with the
observed spectrum of SN 1994D. Clearly the synthetic spectra bear
little resemblance to the observed SN~Ia spectrum.  The \ion{Si}{2}
$\lambda6355$ line is quite weak and may not extend to high enough
velocity. This is a generic problem with sub-Chandrasekhar
models. The 15-d WW2 model of Figure~\ref{woos1} does however reproduce the
boxy shape of the Ca IR triplet seen in SN 1994D, which may indicate
that the calcium is confined to the correct velocity range in the model.

\begin{figure}
\epsscale{0.8}
\plotone{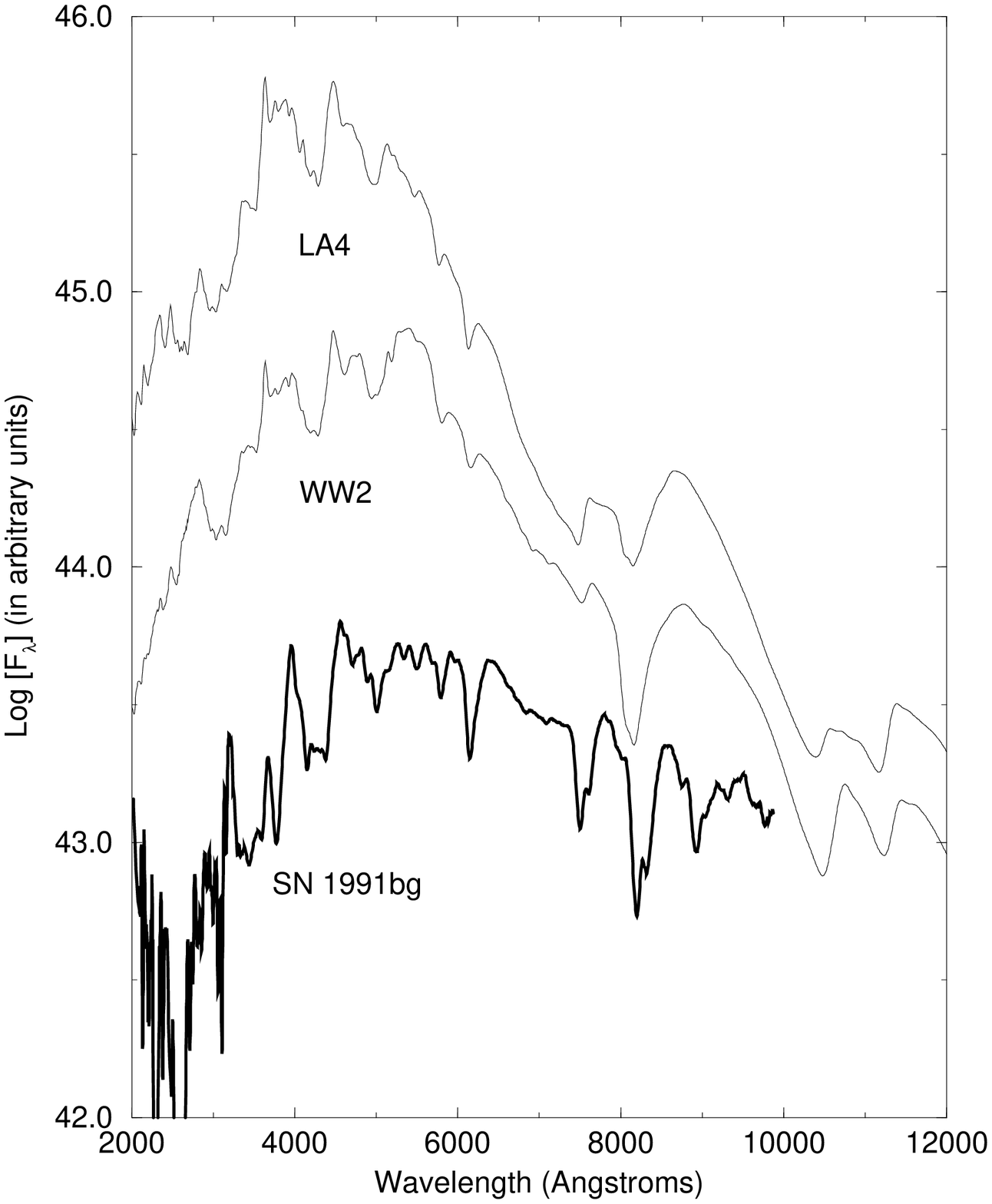}
\caption{\label{subc_comp2}The synthetic spectra of WW2 ($\eta=1.1$)
and LA4 ($\eta=1.5$) at 20~d compared with the observed spectrum of SN
1991bg [at maximum light \protect\cite[]{fil91bg92,iue91bg}].}
\end{figure}

Sub-Chandrasekhar models have been suggested as attractive models for
low luminosity SNe~Ia such as SN~1991bg. The spectra of this SN Ia
[at maximum \cite[]{fil91bg92} in the optical and IUE data in the UV
\cite[]{iue91bg}] along with the spectra of the 20~d 
models of WW2 ($\eta=1.1$) and LA4 ($\eta=1.5$) can be seen in
Fig.~\ref{subc_comp2}.  These models show some resemblance to the
observed SN~1991bg spectrum, however, the shape of the spectra from
these models is counter to that observed. This behavior is exemplified
by the WW2 20~d model with $\eta=1.1$ in Table~\ref{tab2} where $B-V$
is extremely red (0.74), but $U-B$ is negative (-0.16). The flat
spectrum in the UV for these models (signified by a negative $U-B$) is
observationally associated with bright SNe~Ia such as SN~1991T and
SN~1994D, rather than with dim supernovae such as SN~1991bg. In the
spectrum of SN~1991bg, the trough near 4000~\ang\ is 
due to \ion{Ti}{2} \cite[]{fil91bg92,nugseq95}. In the synthetic spectra
this trough is not flat enough to reproduce the observed spectrum of
SN~1991bg, most likely because \ion{Ti}{2} is confined to a small region in
velocity in these models. Also, the calculated \ion{Si}{2} and
\ion{Ca}{2} lines 
are very weak even at this epoch. The strength of these lines
correlates inversely with the luminosity of the supernova
\cite[]{nugseq95}.  Weak, dim supernova like SN~1991bg have strong
\ion{Si}{2} and \ion{Ca}{2} lines, whereas powerful, bright SNe such
as 1991T, have weak lines, thus a model that hopes to fit SN~1991bg
should show prominent \ion{Si}{2} and \ion{Ca}{2} lines.

Table~\ref{tab3} lists the relative concentration of the four most
abundant species at three different velocities for each of the three
models that we have examined (the models W7, WW2, and LA4 have
$\eta=1.0, 1.0, 1.5$, respectively) at 20 days after explosion. The
models have very similar compositions at low velocity, a mixture of
highly ionized iron-peak elements and \ion{He}{2}. The detonation
models have somewhat more \ion{He}{2} than does the deflagration
W7. In the outer parts, the composition is different. Whereas W7 has
singly and doubly ionized intermediate mass elements, the helium
igniters have mixtures of neutral and singly ionized helium, and
doubly ionized elements just below the iron-peak (calcium and
titanium).  At very high velocity ($v > 20000$~\kmps), the composition
of W7 is mostly a mixture of C--O, while the WW2, and LA4 models are
dominated by \ion{He}{1}~--~II. As we have already noted the presence
of intermediate mass elements at high velocity is required
to fit the observed spectra.

\begin{figure}
\epsscale{0.8}
\plotone{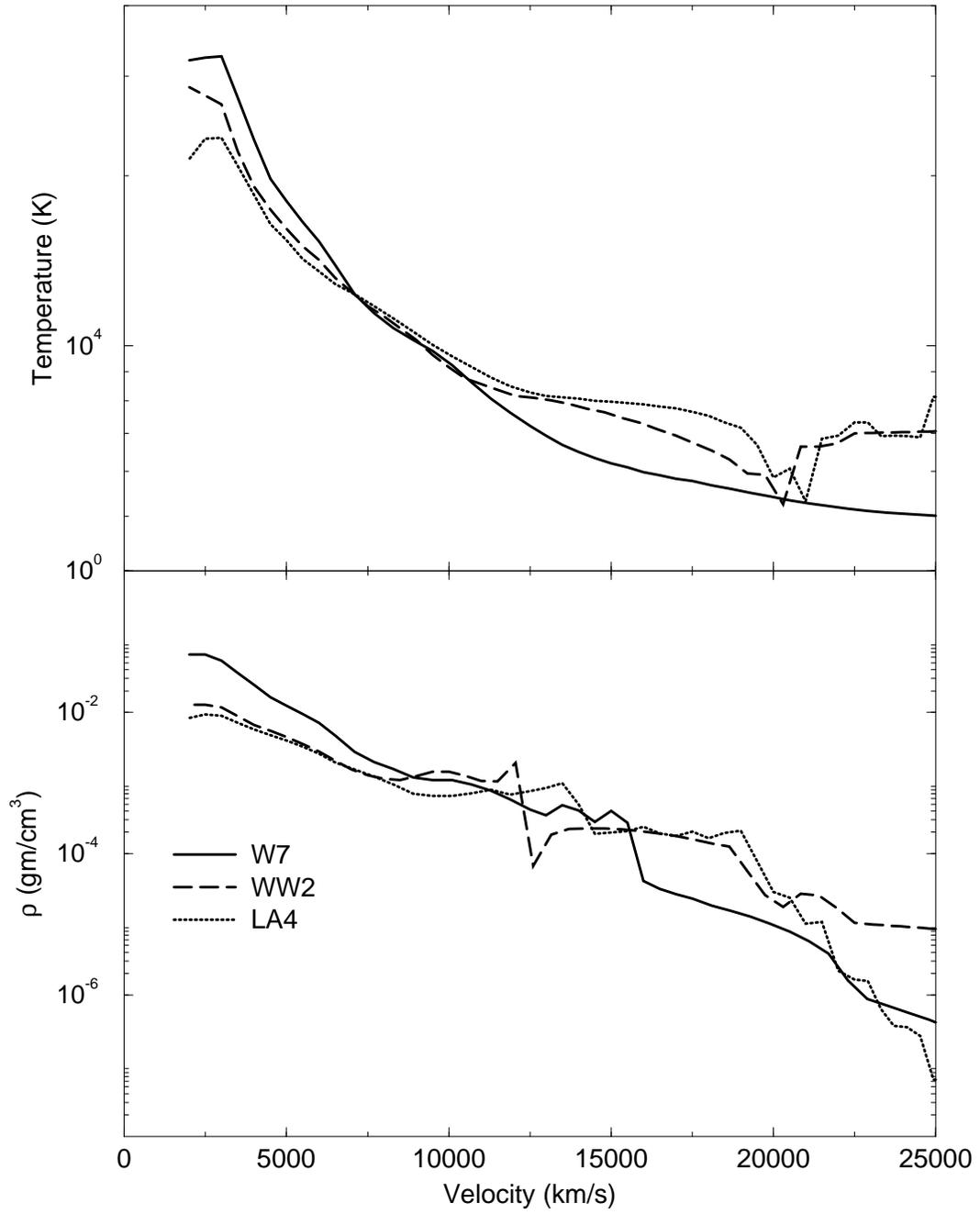}
\caption{\label{temp}The electron density and temperature profiles for
WW2 and W7 at 20~d, with $\eta = 1.1, 1.0$ respectively. The
``glitches'' in the density and temperature occur at the edge of the
burning fronts and represent real changes in model structure.}
\end{figure}

\begin{figure}
\epsscale{0.8}
\plotone{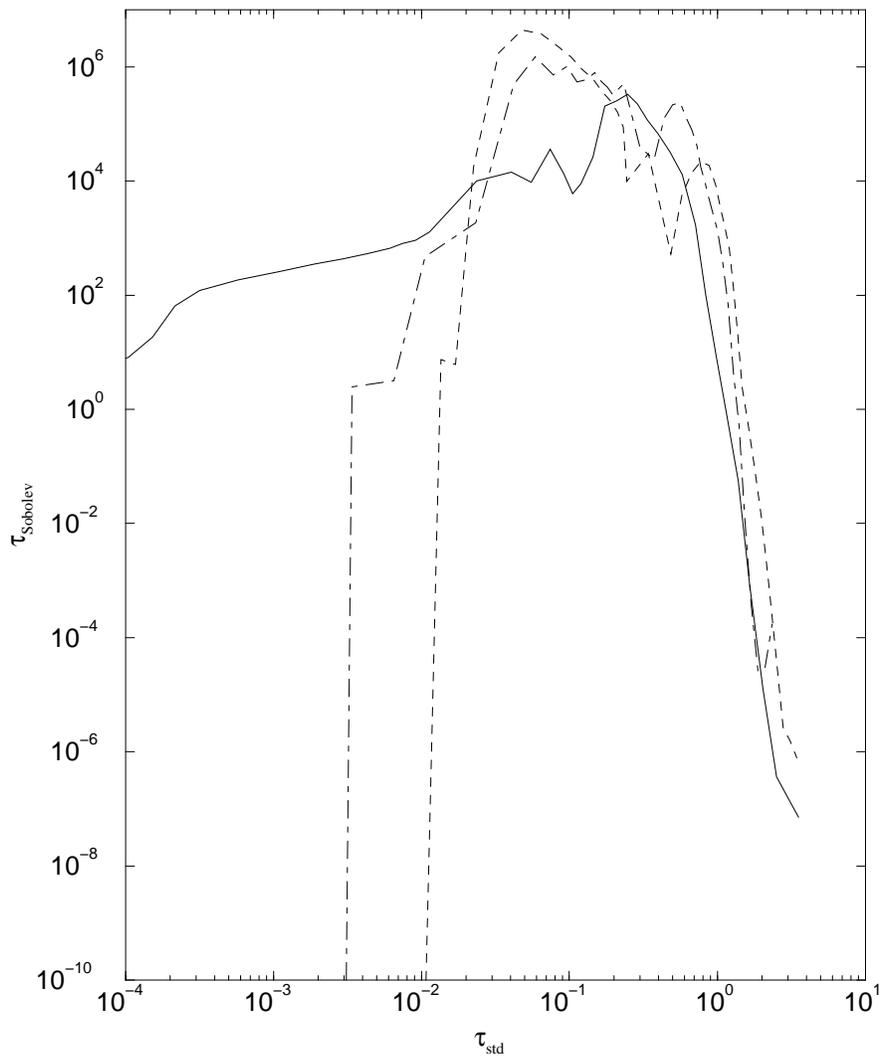}
\caption{\label{depths}The Sobolev optical depth of the Co~II
$^1D-^3G^o\ \lambda 2605.2$ line as a function of \tstd for the 3
models listed in Table~\protect\ref{tab3}. The solid line denotes the
W7 model, the dashed line WW2, and the dot-dashed line LA4. Note that
$\tau_{\rm Sobolev}$ is a purely local quantity and hence is not necessarily
monotonic.}
\end{figure}

In order to further elucidate the differences between W7 and the
helium-igniters, Figure~\ref{temp} compares the temperature and
electron density profiles of W7 ($\eta=1.0$) to WW2 ($\eta=1.1$) at
20~d. The more massive W7 model has a much steeper electron density
profile, and it is cooler on the outside. Figure~\ref{depths} displays
the Sobolev optical depth of the \ion{Co}{2} $^1D-^3G^o\ \lambda 2605.2$
line as a function of \tstd for the 3 models. While W7 and WW2 display
similar Sobolev optical depths in this line at depth, LA4 only has a
very small region where this line is optically thick, and, near the
surface where the spectrum forms, both WW2 and LA4 are transparent in
this line. This is typical for the iron-peak UV lines and it makes it
very difficult for the helium-detonation models to display the proper
line blanketing in the UV. It is not that the iron-peak elements
are not present, but rather that they are not in the proper
ionization/electronic states to create strong line blanketing.

\section{Conclusions} 

We have calculated very detailed NLTE synthetic spectra of
hydrodynamical models for SNe~Ia. We have used only symmetry
considerations at the inner boundary and thus have not had to make any
assumptions about the form of the flux there, the spectrum is
calculated {\em ab-initio}. We have used a simple but accurate
$\gamma$-ray transport algorithm and we have developed a reasonable
parameterization of the time dependence of the $\gamma$-ray heating
that can be compared with and calibrated to sophisticated radiation
hydrodynamical calculations as they become available.

The Chandrasekhar-mass deflagration model W7 shows good agreement with
observed normal SNe~Ia and it is likely that other Chandrasekhar-mass
models such as DD or PDD \cite[]{kmh93} will also show reasonable
agreement. While the sub-Chandrasekhar mass ``helium igniter'' models
bear some resemblance to the sub-luminous SNe~Ia typified by SN~1991bg
the weakness of lines of the intermediate mass elements and the lack
of the UV deficit will have to be addressed if these models are to
remain as viable contenders for at least some SNe~Ia.

\acknowledgements We thank Dave Arnett, Ken Nomoto, and Stan Woosley
for providing us with their models and for helpful discussions. We
also thank Shiomi Kumagai and Tim Young for allowing us to quote their
unpublished results and for helpful discussions on gamma-ray
deposition. This work was supported in part by NSF grants AST-9417242
and AST-9417102; an IBM SUR grant to the University of Oklahoma; and
by NASA grants NAGW-4510, NAGW-2628, and NAGW 5-3067 to Arizona State
University.  Some of the calculations presented in this paper were
performed at the Cornell Theory Center (CTC), and the San Diego
Supercomputer Center (SDSC), supported by the NSF, and at the National
Energy Research Supercomputer Center (NERSC), supported by the
U.S. DoE. We thank all of these institutions for a generous allocation
of computer time.

\bibliography{refs,sn1a,crossrefs}

\clearpage

\begin{deluxetable}{crrrrrr}
\tablecaption{Magnitudes and Colors of W7 Models \label{tab1}}
\tablehead{ \colhead{Time} & \colhead{$\eta$} &
\colhead{$M_{\rm Bol}$} & \colhead{$M_B$} & \colhead{$M_V$} &
\colhead{$B-V$} & \colhead{$U-B$} } 
\startdata 
16 d & 0.6 & -18.48 & -18.64 & -18.60 & -0.04 & -0.35\\ 
16 d & 0.9 & -18.85 & -18.95 & -18.78 & -0.18 & -0.64\\ 
16 d & 1.1 & -19.10 & -19.11 & -18.92 & -0.19 & -0.79\\
16 d & 1.7 & -19.61 & -19.44 & -19.26 & -0.17 & -0.91\\
18 d & 0.9 & -18.70 & -18.93 & -18.83 & -0.10 & -0.37\\
20 d & 0.8 & -18.51 & -18.71 & -18.81 &  0.10 & -0.09\\
20 d & 0.9 & -18.55 & -18.77 & -18.83 &  0.07 & -0.12\\
20 d & 1.0 & -18.69 & -18.93 & -18.93 &  0.00 & -0.17\\
20 d & 2.0 & -19.45 & -19.54 & -19.24 & -0.30 & -0.82\\
23 d & 1.3 & -18.75 & -18.93 & -19.04 &  0.10 & -0.14\\
25 d & 1.5 & -18.81 & -19.11 & -19.08 & -0.04 & -0.20\\
\enddata 
\tablecomments{$M_{\rm Bol}$, $M_B$, and $M_V$ are the bolometric, $B$
and $V$ absolute magnitudes of the models respectively. $B-V$ and
$U-B$ are the associated colors.}
\end{deluxetable}

\clearpage

\begin{deluxetable}{ccrrrrrr}
\tablecaption{Magnitudes and Colors of the Sub-Chandrasekhar Models
\label{tab2}} 
\tablehead{ \colhead{Model} & \colhead{Time} &
\colhead{$\eta$} & \colhead{$M_{\rm Bol}$} & \colhead{$M_B$} &
\colhead{$M_V$} & \colhead{$B-V$} & \colhead{$U-B$} }
\startdata
WW2 & 15 d & 0.7 & -18.08 & -18.04 & -18.68 &  0.64 & -0.18\\
WW2 & 15 d & 1.0 & -18.49 & -18.72 & -18.82 &  0.10 & -0.31\\
WW2 & 15 d & 1.3 & -18.78 & -19.13 & -18.83 & -0.30 & -0.46\\
WW2 & 15 d & 1.6 & -18.98 & -19.33 & -18.85 & -0.48 & -0.55\\
WW2 & 20 d & 1.1 & -18.22 & -18.21 & -18.95 &  0.74 & -0.16\\
WW2 & 20 d & 1.5 & -18.51 & -18.78 & -18.97 &  0.19 & -0.34\\
WW2 & 20 d & 1.9 & -18.80 & -19.17 & -18.82 & -0.36 & -0.65\\
LA4 & 15 d & 0.7 & -18.20 & -18.30 & -18.81 &  0.52 & -0.17\\
LA4 & 15 d & 1.1 & -18.69 & -18.99 & -18.90 & -0.08 & -0.43\\
LA4 & 15 d & 1.7 & -19.17 & -19.41 & -18.91 & -0.50 & -0.76\\
LA4 & 20 d & 1.5 & -18.60 & -18.96 & -18.97 &  0.01 & -0.39\\
LA4 & 20 d & 1.6 & -18.66 & -19.03 & -18.97 & -0.07 & -0.43\\
LA4 & 20 d & 2.0 & -18.89 & -19.29 & -18.91 & -0.37 & -0.60\\

\enddata 
\tablecomments{ $M_{\rm Bol}$, $M_B$, and $M_V$ are the bolometric,
$B$ and $V$ absolute magnitudes of the models respectively. $B-V$ and
$U-B$ are the associated colors.}
\end{deluxetable}

\clearpage

\begin{deluxetable}{lrllll}
\tablecaption{4 Most Abundant Species for Each Model at 20 Days After
Explosion \label{tab3}} 
\small
\tablehead{\colhead{Model} & \colhead{Velocity} &
\multicolumn{4}{c}{Relative Concentration}\\
\colhead{}&\colhead{(km s$^{-1}$)}&\multicolumn{4}{c}{(ppm)}}
\startdata
W7 & 20,000 & \ion{C}{1} ($2\times 10^{5}$) & \ion{O}{2} ($2\times 10^{5}$) &
\ion{Ne}{2} ($2\times 10^{4}$) & \ion{O}{1} ($1\times 10^{4}$)\\ 
LA4 & 20,000 & \ion{He}{2} ($4\times 10^{5}$) & \ion{He}{1} ($9\times 10^{4}$) &
\ion{Ti}{3} ($6\times 10^{3}$) & \ion{Ca}{3} ($4\times 10^{3}$)\\
WW2 & 20,000 & \ion{He}{2} ($4\times 10^{5}$) & \ion{He}{1} ($2\times 10^{5}$) &
\ion{Ti}{3} ($7\times 10^{3}$) & \ion{Ca}{3} ($6\times 10^{3}$)\\
W7 & 10,000 & \ion{Si}{3} ($1\times 10^{5}$) & \ion{Co}{3} ($7\times 10^{4}$) &
\ion{S}{3} ($6\times 10^{4}$) & \ion{Ca}{3} ($3\times 10^{4}$)\\
LA4 & 10,000 & \ion{Si}{3} ($1\times 10^{5}$) & \ion{Co}{3} ($1\times 10^{5}$) &
\ion{S}{3} ($5\times 10^{4}$) & \ion{Ni}{3} ($1\times 10^{4}$)\\
WW2 & 10,000 & \ion{Si}{3} ($2\times 10^{5}$) & \ion{S}{3} ($1\times 10^{5}$) &
\ion{Ar}{2} ($1\times 10^{4}$) & \ion{Ca}{3} ($1\times 10^{4}$)\\
W7 & 5,000 & \ion{Co}{4} ($2\times 10^{5}$) & \ion{Ni}{4} ($7\times 10^{4}$) &
\ion{Fe}{4} ($2\times 10^{4}$) & \ion{He}{2} ($2\times 10^{3}$)\\
LA4 & 5,000 & \ion{Co}{4} ($2\times 10^{5}$) & \ion{He}{2} ($7\times 10^{4}$) &
\ion{Ni}{4} ($2\times 10^{4}$) & \ion{Fe}{4} ($2\times 10^{4}$)\\
WW2 & 5,000 & \ion{Co}{4} ($2\times 10^{5}$) & \ion{He}{2} ($3\times 10^{4}$) &
\ion{Ni}{4} ($2\times 10^{4}$) & \ion{Fe}{4} ($2\times 10^{4}$)\\
\enddata
\tablecomments{The relative concentrations by number of the four most
abundant species (in parts per million) for a particular zone (labeled
by its velocity). The models W7, WW2, and LA4 have $\eta=1.0, 1.1,
1.4$, respectively.}
\end{deluxetable}

\clearpage

\clearpage

\clearpage

\clearpage

\clearpage

\clearpage

\clearpage

\clearpage

\clearpage

\clearpage

\end{document}